# A magnetic phase-transition graphene transistor with tunable spin polarization

*Péter Vancsó[1]\*, Imre Hagymási[2]\* and Levente Tapasztó[1]*

1. Centre for Energy Research, Institute of Technical Physics and Materials Science, 2D Nanoelectronics „Lendület" Research Group, Budapest, Hungary
2. Wigner Research Centre for Physics, Institute of Solid State Physics and Optics, Strongly Correlated Systems ''Lendület'' Research Group, Budapest, Hungary

\* These authors contributed equally to this work.

## Abstract

**Graphene nanoribbons (GNRs) have been proposed as potential building blocks for field effect transistor (FET) devices due to their quantum confinement bandgap. Here, we propose a novel GNR device concept, enabling the control of both charge and spin signals, integrated within the simplest three-terminal device configuration. In a conventional FET device, a gate electrode is employed to tune the Fermi level of the system *in* and *out* of a static bandgap. By contrast, in the switching mechanism proposed here, the applied gate voltage can dynamically open and close an interaction gap, with only a minor shift of the Fermi level. Furthermore, the strong interplay of the band structure and edge spin configuration in zigzag ribbons enables such transistors to carry spin polarized current without employing an external magnetic field or ferromagnetic contacts. Using an experimentally validated theoretical model, we show that such transistors can switch at low voltages and high speed, and the spin polarization of the current can be tuned from 0% to 50% by using the same back gate electrode. Furthermore, such devices are expected to be robust against edge irregularities and can operate at room temperature. Controlling both charge and spin signal within the simplest FET device configuration could open up new routes in data processing with graphene based devices.**

# Introduction

The ability to manipulate electronic and magnetic signals is at the heart of our information and communication technology. Today the information processing relies on our ability to process electric signals, while data storage is built on manipulating magnetic moments. Devices that can provide control over both charge and spin signals could open up new prospects for future electronics. Graphene has emerged as a material with strong potential for both electronic and spintronic applications[1,2]. The presence of a band gap in the electronic band structure is crucial in order to enable digital electronic applications, conventionally based on devices such as field-effect transistors (FETs). Due to the absence of a band gap in 2D graphene sheets, several strategies have been proposed to enable the complete depletion of charge carriers, and therefore a leakage free OFF state of the transistor. Among these approaches the scientific attention quickly turned towards graphene nanoribbons (GNRs), narrow stripes of graphene, as promising candidates for low-dimensional FETs with increased $I_{on}/I_{off}$ ratio[3,4,5]. The operation of early GNR FET devices fabricated by e-beam lithography was dominated by edge roughness and disorder, which considerably reduces their operation speed, as also evidenced by their low mobility[3]. However, with the rise of more precise nanofabrication techniques[6,7,8,9], as well as bottom-up synthesis[10,11,12], such limitations can in principle be surmounted. Nevertheless, the realization of high mobility electronic devices based on semiconductor GNRs have proven particularly challenging. Classically, armchair graphene nanoribbons have been proposed as building blocks of FET devices, as ribbons of this orientation possess the largest quantum confinement band gap. However, for armchair nanoribbons even a single atomic row variation in their width can drastically alter the bandgap[13]. This implies that such devices are particularly sensitive to structural irregularities and disorder along their length. We propose here that by using zigzag-edge GNRs (ZGNRs) this limitation can be overcome.

The semiconducting nature of ZGNRs is related to interaction effects and their spin polarized edge states[14,15]. The semiconducting state of ZGNRs at half filling, with oppositely spin-polarized edges, is robustly predicted by several theoretical models, including density functional theory (DFT)[15,16,17], the mean-field approximation of the Hubbard model[14,18], exact diagonalization and quantum Monte Carlo simulation [19], as well as density-matrix renormalization group algorithm (DMRG)[20,21]. Furthermore the spin polarized edge states, in principle also enable us to gain access to the spin degree of freedom in such ribbons. It has already been shown that the magnetic configuration of ZGNRs can be tuned by using an external magnetic field[22,23,24,25] or ferromagnetic electrodes[26,27,28,29], which can be exploited in various device applications such as spin valves, spin diodes, spin transistor and GMR devices. However, controlling the device by means of external magnetic fields is technologically cumbersome, as compared to a more desirable electrical control. In this direction, Son et al.[30] reported the concept of a half-metallic ZGNR transistors that can carry fully spin polarized current, without the application of a magnetic field, only by using an in-plane external electric field across the ribbon. In this case due to the spatially separated edge states, the transverse electric field closes the gap for one spin orientation and widens it for the other, resulting in half-metallicity. Several calculations were performed using transverse electric fields for possible spintronic applications[31,32,33,34]. However, from the device fabrication point of view the application of a homogeneous in-plane electric field across a few nanometer wide ribbon is technologically quite challenging and implies a large lateral device size, which is more difficult to integrate. It would be a huge technological advantage to control the spin signal in ZGNRs by conventional back/top gate electrodes, involving a perpendicular electric field.

We propose that this can be achieved by exploiting a novel switching mechanism in narrow zigzag GNRs enabled by doping the ribbons (moving away from the half-filling). It is known that slightly above the antiferromagnetic (AFM) ground state energy a ferromagnetic (FM) configuration exists with parallel spin orientation on both edges[16]. In contrast to the half-filled system, where the AFM state is always more stable, in a doped ZGNR, both AFM and FM sates can become energetically favorable. Most importantly, there is an intriguing interconnection of the spin configuration and the electronic band structure that renders the ribbons in the AFM spin configuration insulating[15], while FM nanoribbons are metallic[17]. We have recently shown experimentally that a sharp transition from semiconducting (AFM) to metallic (FM) state can indeed occur as a function of the ribbon width[35]. However, varying the ribbon width is not a practical parameter for device applications. In this study, by using the same experimentally validated theoretical model, we show that the sharp semiconductor (AFM)-metal (FM) transition can also be induced through changing the carrier density ($n_{2D}$) of the ZGNR[18,36,37]. Our calculations, based on the Hubbard model and the Landauer transport formalism, reveal that the novel switching mechanism based on magnetic phase transition enables excellent FET characteristics, and the device can also carry spin-polarized current. This is enabled by the fact that in contrast to the conventional FETs, where upon gating the electronic structure remains unchanged, only the charge density is modulated, in an interacting system the density of states depends strongly on the electron density. The main advantage of our device concept is that the carrier density in ZGNR can be easily tuned using a back gate electrode, therefore neither external magnetic field nor transverse electric field is required to control both the charge and spin currents in the device.

**Results and discussion**

Figure 1 shows the schematic idea of the proposed device concept, comprising two nonmagnetic metal contacts and a back-gate electrode. A long, narrow, zigzag graphene nanoribbon of the width $w$ is placed on a gate insulator which has a thickness $d$ and a relative dielectric constant $\varepsilon_r$.

First, we calculate the electronic and magnetic properties of the ZGNR constituting the channel material of the device using the mean-field approximation of the Hubbard-model accounting for both the doping of the ribbon and the effect of finite temperature:

$$\mathcal{H} = -\sum_{\langle i,j \rangle} t_{ij} \hat{c}_{i\sigma}^\dagger \hat{c}_{j\sigma} + U \sum_j \hat{n}_{j\uparrow} \hat{n}_{j\downarrow} - \mu \hat{N}.$$

The first term is the nearest-neighbor tight-binding Hamiltonian, while the second term accounts for the onsite Coulomb repulsion. The effects of temperature and doping have been included in the third term through the grand canonical ensemble, where $\mu$ is the chemical potential, and $\hat{N}$ the particle number operator. The electron density and the chemical potential were determined self-consistently with the interaction parameter U = 3.24 eV estimated from experiments[35] (see Supplementary Information). In the calculations we considered hydrogen-saturated and relaxed edges similar to first-principles DFTcalculations[15]. A nanoribbon is a quasi-one-dimensional system, where quantum fluctuations can be significant; hence, it is not straightforward that mean-field theory can give an accurate description of the system. However, for wide enough ribbons – like in our case – it has been shown based on QMC results[38] that mean-field theory gives quantitatively good description in the weakly interacting limit, U < 5.4 eV.

Using the above theoretical model, in a previous work, we were able to quantitatively interpret the sharp semiconductor-metal transition experimentally observed as a function of ribbon width in zigzag graphene nanoribbons defined by STM Lithography[35]. Now, using the same theoretical model, we found that the semiconductor (AFM)-metal (FM) phase transition can also occur as a function of charge carrier concentration. This finding is of huge practical importance, as the charge carrier concentration can be easily tuned, in contrast to the ribbon width, enabling us to exploit the opening of an interaction gap associated with the phase transition in device applications. Figure 2a shows the phase diagram of ZGNRs as a function of ribbon width and charge carrier density, as calculated from the Hubbard Hamiltonian above. Due to the fact that our lattices are bipartite, particle-hole symmetry is present; therefore, our results are valid both for the n-type or p-type doping. The most important finding is that for realistic ribbon widths of orders of nanometers, the phase transition occurs in the low doping regime ($n_{2D} \sim 10^{12}$ cm$^{-2}$) which can be easily accessed experimentally by simple electrostatic gating. Concerning the magnetic properties of such ZGNRs, in the low doping regime we have found only two types of magnetic phases, AFM and FM, which are the result of the competition between band and interaction energies. The effect of doping can be understood by considering that the extra electrons start occupying states of the conduction band minima. However, in the AFM configuration, due to the presence of a significant band gap, this is energetically quite unfavorable. Consequently, when the doping reaches a critical value, it is more favorable for the system to switch to a metallic state and gain band energy, by changing the spin configuration on the edges from AFM to FM. As can be seen from the phase diagram in Fig 2a, the critical doping value depends on the ribbon width and the temperature. By decreasing the width of the ribbons the AFM/FM switching occurs at higher doping levels. Regarding the temperature, the AFM-FM phase transition is shifted towards the higher doping values as the temperature is increased. For example in a 20-ZGNRs (ZGNR that is twenty carbon zigzag lines wide, $w$= 4.2 nm) the critical doping value is $1.6 \times 10^{12}$ cm$^{-2}$ at $T$=60K compared to $2.6 \times 10^{12}$ cm$^{-2}$ at $T$=300K, which means that the AFM ground state is realized in a wider parameter range at higher temperatures. In Figure 2b we plot the band gap values ($E_g$) and the total magnetic moment ($M = 1/2(n_\uparrow - n_\downarrow) g_e\mu_B$) per unit length ($a_0$=0.246 nm) as a function of the doping for a 20-ZGNR at $T$=300K. In the AFM configuration the band gaps were found to be about $E_g \sim 0.25\ eV$ with zero ($M$=0) total magnetic moment resulting from the opposite spin directions at the edges. By contrast, in the FM configuration the emerging net magnetic moment is about $M \sim 0.6\ \mu_B/a_o$ (Bohr magnetron per unit length), with no band gap ($E_g$=0 $eV$). The band structures shown in the insets of Fig 2b correspond to the $n_{2D}$=1×10$^{12}$ cm$^{-2}$ (AFM) and $n_{2D}$=4×10$^{12}$ cm$^{-2}$ (FM) case, respectively. We found that in the low doping regime the band gaps and magnetic moments display only a small variation as a function of carrier density, except near the transition point. It is also worth mentioning that in the higher doping regime further magnetic phases appear[18] until the edge-state bands are filled and the ribbon becomes paramagnetic (near 10$^{14}$ cm$^{-2}$ doping values)[18,39]. However, such doping levels are difficult to achieve in a solid gate device configuration.

To calculate the device characteristics besides the properties of the ZGNRs, the gate capacitance ($C = e\partial n_{2D}/\partial V_G$), which is the relationship between the applied gate voltage ($V_G$) and the induced carrier density ($n_{2D}$) at room temperature, has also to be accurately determined. Our detailed calculations reveal that besides the classical capacitance ($C_C$), the quantum contribution ($C_Q$) is also significant, as it is often the case in nanoscale devices[40] (for details see Supplementary Information).

The quantum capacitance of the ZGNRs was numerically derived from our Hubbard calculations, while the classical capacitance of a $w$ width ZGNR was calculated from classical electrostatics[41,42] leading to the following expression:

$$C_C = \varepsilon_0 \varepsilon_r \pi \left[ 2d \arctan\left(\frac{w}{4d}\right) + \frac{w}{4} \ln\left(1 + \frac{16d^2}{w^2}\right) \right]^{-1},$$

where $d$ and $\varepsilon_r$ are the thickness and the relative dielectric constant of the insulator. This analytical expression for $C_C$ is valid as long as the surface charge density of the ZGNR is roughly constant, which is fulfilled in our low doping regime calculations. In addition, this expression describes quite accurately the experimental results of back gated graphene nanoribbons[43]. Figure 3 shows the calculated $n_{2D}$ as a function of $V_G$ for different oxides SiO$_2$ ($\varepsilon_r$=3.9) and HfO$_2$ ($\varepsilon_r$=47)[44] in the case of 20-ZGNR at $T$=300K. Due to our device geometry, where the ZGNR is interfaced by the gate insulator and air, we used an effective dielectric constant in the calculations, which is the average value of the dielectric constants of substrate and air. For a SiO$_2$ insulating layer with d=300 $nm$ and d=30 $nm$ we found that the AFM-FM transition ($n_{2D}$=2.6×10$^{12}$ cm$^{-2}$) occurs at $V_G$=0.96 $V$ and $V_G$=0.67$V$, respectively. As for both layer thicknesses $C_C \ll C_Q$, a linear behavior is obtained as a function of $n_{2D}(V_G)$. In contrast, $C_C \gg C_Q$ for a HfO$_2$ insulating layer with d=30 $nm$, therefore the nonlinear behavior of $V_Q$ (see Figure S1) dominates in the $n_{2D}(V_G)$ function. The phase transition for HfO$_2$ layer was found to occur at $V_G$=0.17 $V$. One can also note that the values of the $n_{2D}$ jump at the critical gate voltage value where the first order magnetic phase transition takes place. From practical point of view, the most important finding is that such devices can already switch at very low gate voltages. Furthermore, the critical value of the gate voltage, at which the AFM-FM transition occurs, can be further decreased by reducing the thickness of the oxide-layer or increasing the dielectric constant. By changing the width of the ZGNR we found that the critical value of the gate voltage further decreases when $w$ is increased as also apparent from Figure 2a, where wider ribbons are characterized by lower critical doping values. Overall, we can conclude that ZGNR transistors can switch in the range of $V_G \approx 0.1$-$1\ V$, depending on the width of the ZGNR and the properties of the gate insulator layer.

To explore the transport characteristics of the proposed novel device, featuring a magnetic phase transition and interaction band gap, the interaction effects need to be incorporated into the transport calculations. This is not straightforward, as the widely used Landauer formalism was developed for non-interacting Hamiltonians. However, by using the mean-field approximation, the original many-body problem can be mapped to an effective single-particle Hamiltonian to which the Landauer's method can be applied[45]. We have included into our calculations the effect of electron interactions by using the spin dependent Hamiltonian matrix elements from our Hubbard calculations. The conductance calculations between the source and drain contacts have been performed based on the Landauer formalism using the standard recursive Green's function technique:

$$G^\sigma = -\frac{e^2}{h}\int dE \sum_{ij} T_{ij}^\sigma(E) \frac{\partial f(E-E_F)}{\partial E},$$

where $\sigma = \uparrow, \downarrow$ is the spin index, $T_{ij}(E)$ is the transmission coefficient from sub-band $j$ in the source to sub-band $i$ in the drain electrode at energy E, while $f$ is the Fermi function. Figure 4 shows the spin dependent conductance $G^\sigma$ as a function of charge carrier density $n_{2D}$ for the defect-free 20-ZGNR at $T$=0K and $T$=300K.

At T=0K (Figure 4a) below the phase transition carrier concentration the conductance is zero according to the presence of the band gap in the AFM configuration (Figure 4c). Reaching the critical value of doping ($n_{2D}$=1.6×10$^{12}$ cm$^{-2}$) the ZGNR becomes metallic (FM configuration) due to the open conduction channels for both up and down spins at the Fermi energy (Figure 4d). The equal number of conduction channels for the different spin directions implies that the current is not spin polarized even though the ribbon itself is spin polarized ($n_\uparrow > n_\downarrow$). However, the current can become spin polarized by further increasing the charge carrier concentration (gate voltage) to $n_{2D}$=8.7×10$^{12}$ cm$^{-2}$ where additional conductance channels open for spin down while the number of conductance channel for spin up state remains the same. This asymmetry of the band structure for up and down spins in the FM configuration (Figure 4e) enables our device to carry spin polarized current. Consequently, our device layout does not require the use of an external magnetic field, ferromagnetic contacts or transvers electric fields to achieve spin polarized current, which is a huge technological advantage. Another intriguing property of our device is that the charge and the spin currents can be independently tuned by the charge carrier density (sweeping the gate voltage). This is due to the somewhat counter-intuitive fact that after switching the ribbon into the FM phase, although the spin polarization (net magnetic moment) of the ribbon becomes nonzero, the spin polarization of the current still remains zero. This can be understood in the Landauer formalism, where the product of electron velocity ($v \sim \partial Ek/\partial k$) and density ($n \sim \partial k/\partial E$) in the current ($I = env$) cancels the $\partial E/\partial k$ term, thus the conductance is determined only by the number of the open conductance channels, independently of the shape of the dispersion curve.

In Figure 4b we plot the conductance of the same 20-ZGNR at room temperature (T=300K). The effect of the temperature has two important effects: (1) it shifts the position of the AFM-FM (semiconductor-metal) transition, and (2) it modifies the shape of the conductance plateaus due to the broadened energy distributions of the incident electrons described by the Fermi-function. In contrast to the step-like transition of both charge and spin current for the T=0 K case, at room temperature both transitions become gradual, due to combined effect of the temperature and doping. In contrast to the sharp switching of spin polarization from 0% to 50% at low temperatures (0K), at room temperature the spin polarization degree of the current can be continuously tuned between 0 and 50% by sweeping the back-gate voltage. This provides us a fully electrical control through a simple back gate electrode over the spin polarization degree of the current, which could be exploited in future spintronic devices. In principle, such a ribbon could act as spin source replacing ferromagnetic electrodes, which would be a huge technological advantage and on the long run enabling all-graphene spintronic circuits[2].

Beside the tunable spin polarization, the FET characteristics of the device are also interesting as it employs a novel switching mechanism. Instead of tuning the Fermi level *in* and *out* of the static bandgap of a channel material, we can dynamically open and close an interaction gap, with only a minor shift of the Fermi level. The calculated $I_{on}/I_{off}$ ratio of a 20-ZGNR device is about ~$10^5$ at T=150K, but decreases to around ~$10^2$ at room temperature. However, this value at room temperature can be significantly increased by using narrower ZGNR characterized by larger band gaps. For instance already a 10-ZGNR ($w = 2.1$ nm) has an $I_{on}/I_{off}$ ratio of about $10^4$ at room temperature, while this value reaches $10^7$ at T=150K. To characterize the switching speed of such a device, we have also calculated its subthreshold swing $S = dV_G/(d(\log I_d))$. The $S$ value is defined to be the amount of change in gate voltage ($V_g$) required to produce a 10x change in drain current ($I_d$) in the low $V_G$ region, where the drain current has an exponential dependence. From the calculated gate voltage and conductance values we can estimate the subthreshold swing of our 20-ZGNR device to $S$ =81

$mv/dec$ in the case of 30 $nm$ HfO$_2$ layer at room temperature. Choosing narrower ribbons (10-ZGNR) this value is further decreased to $S = 71\ mv/dec$. The low $S$ values are comparable with recently reported MOSFETs ($S = 77\ mv/dec$)[46] and single layer MoS$_2$ transistors ($S = 74\ mv/dec$)[47] approaching the fundamental thermionic limitation of the subthreshold swing[48] ($S_{min} = 60\ mv/dec$) in conventional FETs at room temperature. Finally, we note that the large I$_{on}$/I$_{off}$ ratio and low subthreshold swing values of our device are similar compared to previously reported GNR FETs based on more complex structures including both armchair- and zigzag-edge segments[49].

The above results clearly show that both the electronic and spintronic properties of such devices are predicted to be exceptional. However, from practical consideration it is very important to assess how robust such properties are against deviations from the ideal conditions. So far our calculations have been performed on ZGNRs with perfect edges. In reality, several types of deviations from the ideal zigzag edge structure can occur[50,51,52]. Since the operation mechanism of our device is based on the edge magnetism, edge defects and reconstructions that can completely suppress the spins at the edges, in principle, render our device inoperable. Nevertheless, the experimental observation of the AFM/FM transition[35] evidences that such defects are not commonly encountered. STM lithography[7,35] is an effective nanofabrication technique to define graphene nanoribbons with nanometer precision and well-defined crystallographic edge orientations. Therefore we have considered the most likely edge irregularities resulting from this process, namely, atomic scale step-like defects (Figure 5a inset), along an overall zigzag edge orientation. We define the defect concentration ($x/d$) as the ratio of the length of the protrusion ($x=2\cdot a_0$) and the length of the unit cell ($d=n\cdot a_0$), thus Figure 5a inset shows the geometry of 20-ZNGR with 40% edge defect concentration ($n=5$). We found that up to such high edge defect concentrations the ribbon edges still become spin polarized in the half-filled case, although the magnitude of magnetic moment at the edges significantly decreases ($M_{edge}$=0.069 μ$_B$/a$_o$) as compared to the defect-free 20-ZGNR ($M_{edge}$=0.282 μ$_B$/a$_o$) (Fig 5a). We note that a highly similar decrease of the edge magnetic moments was previously predicted for simple edge vacancies and impurities in GNRs[53]. The band structure (Fig 5b) and transport calculations on the highly defective 20-ZGNR at room temperature confirmed the persistence of the semiconductor-to-metal transition as a function of carrier density, and the current flowing through the device becomes spin polarized in a similar manner as for the defect-free ZGNRs. Consequently, the spin-related device properties are predicted to be particularly robust against atomic scale edge disorder, as they persist up to extremely high edge defect concentrations of 40%. It is also very important to note, that the change in the ribbon width introduced by the edge defects only slightly changes the bandgap of such a ZGNR. This is in striking contrast with the case of armchair ribbons, where one atomic wide edge steps can induce an order of magnitude change in the band gap along the ribbon[13]. Consequently, not only the spin but also the electronic transport properties of ZGNRs are expected to be much more robust against edge disorder as compared to their armchair counterparts.

## Conclusion

In summary, we have identified a novel switching mechanism in FET devices based on zigzag graphene nanoribbons, and demonstrated its feasibility to control both charge and spin currents by using a simple gate electrode. We have shown that such devices can switch a low gate voltages (0.1 V - 1 V), and high speed (S = 71 $mv/dec.$), while the spin polarization of the current can be tuned between 0 and 50% by the same gate electrode. We have also shown that the proposed electronic and spintronic device characteristics are particularly robust against edge disorder rendering their operation feasible under realistic conditions even at room temperature. We believe that the device concept presented here allows the simple realization of graphene circuits, enabling the integrated control of both charge and spin signals within the same device that could open new routes for novel data processing techniques.

**Acknowledgements:** The work has been supported by the ERC StG and the "Lendület" programme of the Hungarian Academy of Sciences grants LP2014-14, as well as the Korea Hungary Joint Laboratory for Nanosciences. L.T. acknowledges OTKA grant K10875. I.H. was supported in part by the Hungarian Research Fund (OTKA) through Grant K120569.

**Author contributions:** L.T. proposed the device concept. I.H. and P.V. performed the theoretical calculations. V.P. and L.T. wrote the paper. All authors discussed the results and commented on the manuscript.

# Figures

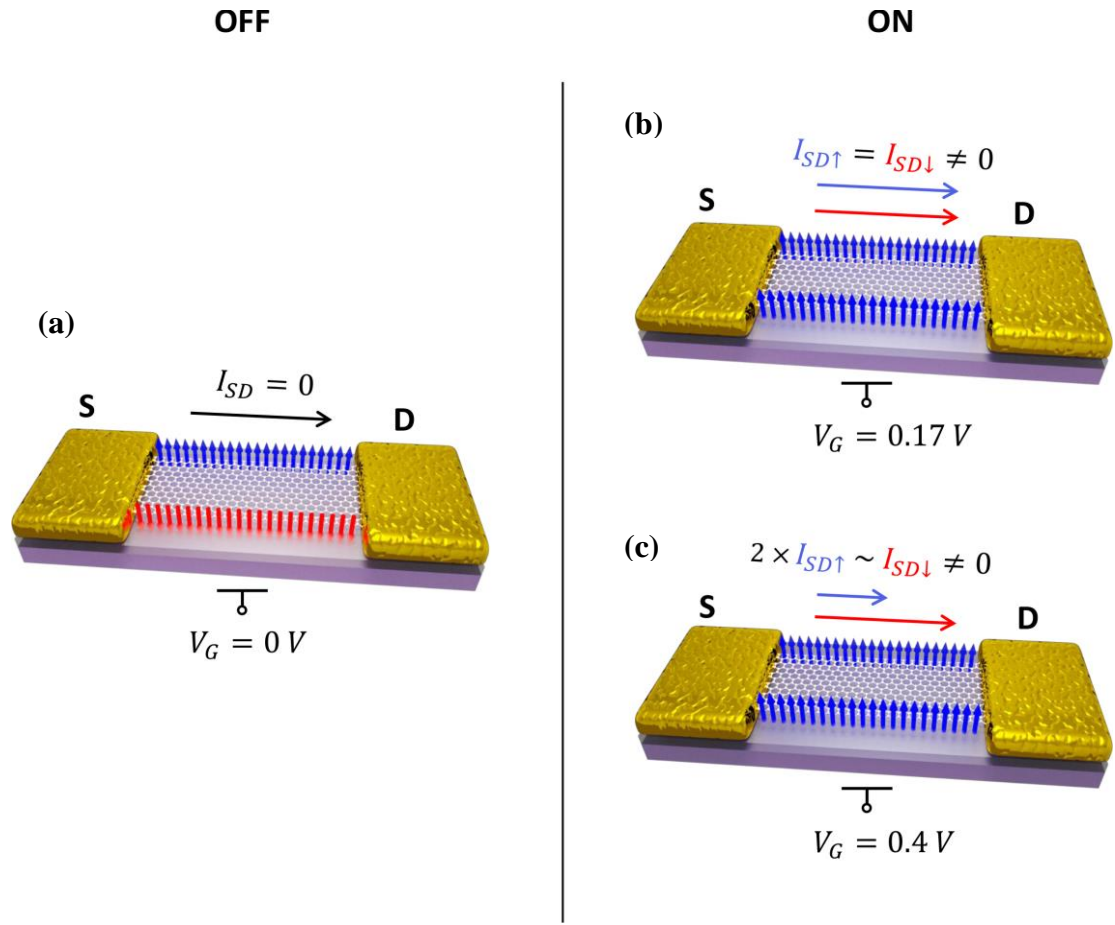

**Figure 1**. Schematic representation of the three states of the proposed ZGNR FET: (a) an insulating OFF state with antiparallel spin orientation on opposite ribbon edges in the absence of a gate bias, (b) the conducting ON state characterized by parallel spin orientation on ribbon edges but no net spin polarization of the current induced by an applied gate voltage, and (c) spin polarized ON state achieved by further increasing the gate potential. ($V_G$ values correspond to 30 $nm$ HfO$_2$ oxide layer at room temperature.)

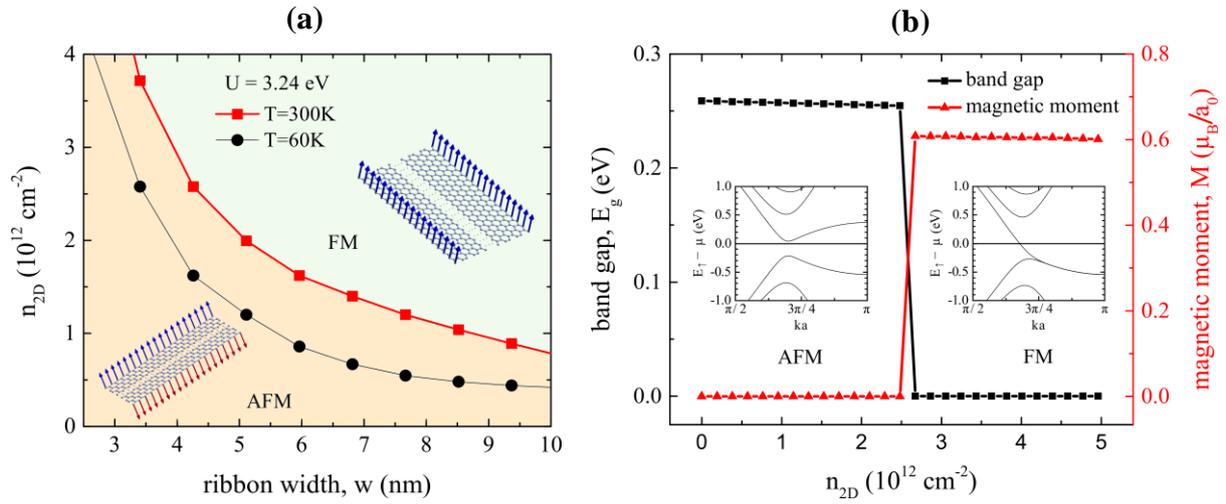

**Figure 2**. (a) Phase diagram of ZGNRs as a function of charge carrier concentration ($n_{2D}$) and ribbon width ($w$) at two different temperatures (T=60K and T=300K). The AFM ground state is the only stable without external doping; however, the FM configuration can become favorable above a critical doping value and ribbon width. (b) A sharp transition in the band gap and magnetic moment of a ZGNR ($w$=4.2 nm) occurs as a function of charge carrier density at room temperature. Insets correspond to the band structures for one spin orientation of the 20-ZGNR at $n_{2D}$=1×10$^{12}$ cm$^{-2}$ (AFM state) and at $n_{2D}$=4×10$^{12}$ cm$^{-2}$ (FM state).

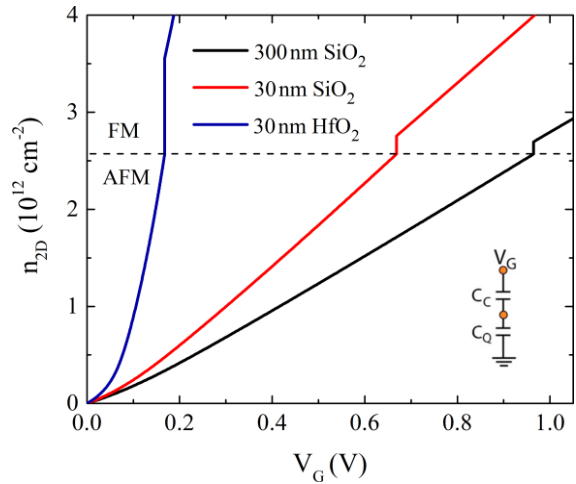

**Figure 3**. Carrier density ($n_{2D}$) in 20-ZGNR ($w$=4.2 nm) as a function of gate voltage ($V_G$) for different oxide thicknesses at room temperature. Such a device can switch at a critical charge density marked by dashed line, corresponding to very low gate voltages.

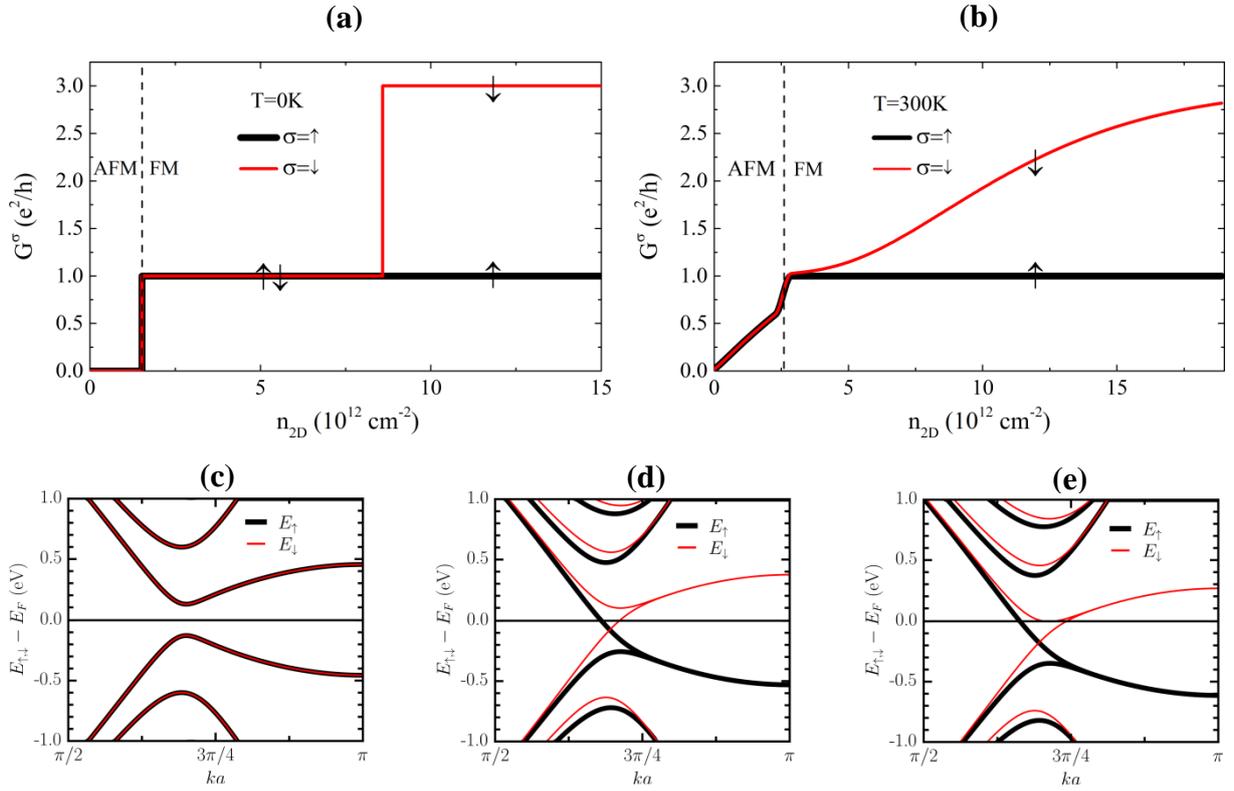

**Figure 4**. Spin dependent conductance ($G^\sigma$) as a function of charge carrier density ($n_{2D}$) for a 20-ZGNR at (a) T=0K and (b) T=300K. The AFM-FM transition (dashed line) dynamically closes the bandgap of the ribbon switching the transistor into a conducting state. Further increasing the carrier concentration opens additional spin down channels leading to the spin polarization of the current. (c-e) Spin resolved band structure in the three operation regimes for: 0 cm$^{-2}$, 3×10$^{12}$ cm$^{-2}$, and 10$^{13}$ cm$^{-2}$ at T=0K, respectively.

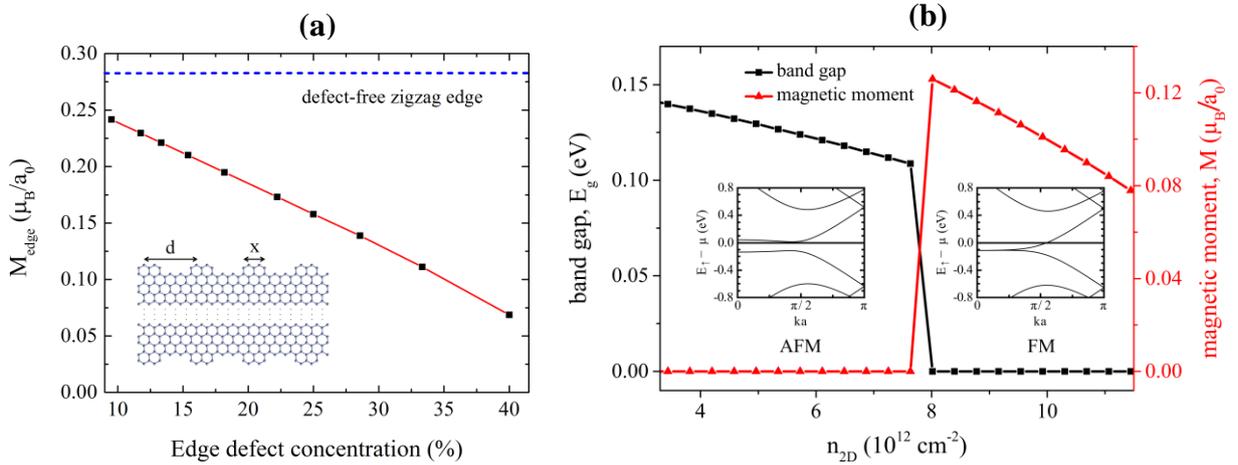

**Figure 5** (a) Edge magnetic moments ($M_{edge}$) as a function of edge defect concentration for 20-ZGNR. Inset shows the geometry of 20-ZNGR with 40% edge defect concentration. Dashed line corresponds to edge magnetic moment of the defect-free 20-ZGNR. (b) Band gap and magnetic moment of highly defective (40%) 20-ZGNR at room temperature as a function of doping. Insets correspond to the band structures for one spin orientation in the AFM ($n_{2D}$=5×10$^{12}$ cm$^{-2}$) and FM ($n_{2D}$=9×10$^{12}$ cm$^{-2}$) state, respectively.